\documentclass[twocolumn,prl,aps]{revtex4}
\usepackage{amsmath}   
\usepackage{amssymb}
\usepackage{amsfonts}
\usepackage{graphicx}

\begin{document}
\title
{Drastic Reduction of Plasmon Damping in Two-Dimensional Electron Disks}

\author{P.~A.~Gusikhin, V.~M.~Muravev,  A.~A.~Zagitova, I.~V.~Kukushkin}
\affiliation{Institute of Solid State Physics, RAS, Chernogolovka, 142432 Russia}
\date{\today}

\date{\today}

\begin{abstract}
The plasmon damping has been investigated using resonant microwave absorption of two-dimensional electrons in disks with different diameters. We have found an unexpected drastic reduction of the plasmon damping in the regime of strong retardation. This finding implies large delocalization of retarded plasmon field outside the plane of the two-dimensional electron system. A universal relation between the damping of plasmon polariton waves and retardation parameter is reported.   
\end{abstract}

\pacs{73.23.-b, 73.63.Hs, 72.20.My, 73.50.Mx}
\maketitle

Studies of plasma excitations in 2DESs have attracted an increasing interest in recent years~\cite{Heitmann:86, Lusakowski:16}. This interest has been triggered by a discovery of an absolutely new part of 2D plasmon spectrum, where coupling with light is especially strong~\cite{Kukushkin:03, Muravev:06}. This regime, where the influence of the retardation is important, can be realized when the plasmon and light wavelengths become comparable. The influence of the electrodynamic effects can be quantified by the retardation parameter $A$, defined as the ratio of the frequency of the quasi-static 2D plasmon $\omega_p$ to the frequency of light in the medium $\omega_{\rm light} = c q /\sqrt{\varepsilon}$ with the same wave vector $q$~\cite{Stern:67}  
\begin{equation}
A=\frac{\omega_p(q)}{\omega_{\rm light} (q)} = \sqrt{\frac{n_s e^2}{2 m^{\ast} \varepsilon_0 \bar{\varepsilon}} q} \times \frac{1}{\omega_{\rm light} (q)} \sim \sqrt{\frac{n_s}{q}},
\label{1}
\end{equation}
where $n_s$ and $m^{\ast}$ are the density and effective mass of the 2D electrons, respectively. Here, effective permittivity of the surrounding medium $\bar{\varepsilon} = (\varepsilon + 1)/2$ is the average dielectric constant of GaAs ($\varepsilon = 12.8$) and free space. These experiments revealed very interesting and fully unexpected features of the 2D magnetoplasmon spectra. It has been shown that in the small wave-vector and high-density regimes, where the influence of the retardation effects is prominent, there is a strong reduction of the resonant plasma frequency with very unusual zigzag behavior in $B$ field.   

These experiments served as a benchmark for a plethora of research in the field of 2D plasma electrodynamics, such as the strong and ultrastrong coupling between 2D electrons and photonic cavity mode~\cite{Gunter:09, Todorov:10, Muravev:11, Science:12, Zhang:16}, super-radiant Dicke decay of the 2D electron ensemble ~\cite{Zudov:14, Sirtori:15, Muravev:16}, and fine structure of the cyclotron resonance~\cite{Muravev:16}. It is widely accepted that retardation is associated with a dramatic broadening of the plasmon polariton resonance~\cite{Sonnichsen:02, Mikhailov:05, Andreev:14}. The main reason for that is a significant increase of radiative decay. This fact has conceptually hampered the progress of research in the direction of plasmonic systems placed in the regime of ultra strong retardation. The regime which is potentially very attractive for plasmonic applications~\cite{Shaner, Knap, Popov}.  

In this Letter we systematically studied the plasmon damping in 2DES disks as a function of their diameter. The measurements reveal that, in contrast to general belief, there is a drastic suppression of plasmon damping in the regime of strong retardation. Our results allowed us to draw important conclusions about delocalization of retarded plasmon field outside the plane of 2DES. Moreover, we established a universal relation between the damping of plasmon polariton waves and retardation parameter. These results pave the way for research in the field of ultra strong light-matter interaction.

This study was performed using a set of high-quality GaAs/Al$_x$Ga$_{1 - x}$As ($x = 0.3$) heterostructures hosting a 2DES in a $(20 - 30)$~nm quantum well. The electron density was in the range of $0.8$ to $6.0 \times10^{11}$~cm$^{-2}$, with an electron transport mobility in the range of $0.5$ to $12\times10^{6}$~cm$^2/$(V$\cdot$~s). Disk-shaped mesas with diameters in the range of $0.05$~mm to $12$~mm were fabricated from these heterostructures. The plasma wave in the disk was excited by electromagnetic radiation guided to the sample either through the waveguide section or with a coaxial cable. The detailed description of the experimental scheme is given in the Supplemental Material~\textrm{I}~\cite{Supplemental}. In the latter case, the radiation from the coaxial cable was transferred to the 2DES through a coplanar waveguide transmission line. In order to detect 2D electron plasma excitations, we employed a non-invasive optical technique of microwave absorption detection~\cite{Ashkinadze, Kukushkin:02}. This technique relies on the high sensitivity of the 2DES luminescence spectra to 2D-electrons heating caused by the absorption of incident microwave radiation. In this technique, 2DES luminescence spectra with and without microwave radiation were collected using a spectrometer with a liquid-nitrogen cooled charge-coupled device (CCD) camera. The resulting differential spectrum reveals microwave-caused 2DES temperature heating. Therefore, the integral of the absolute value of the differential spectrum could be used as a measure of the microwave absorption by the 2DES. In our experiments, we used a $780$~nm stabilized semiconductor laser with a net output power of $4$~mW to excite 2D luminescence, and single-grating spectrometer with a spectral resolution of $0.25$~meV to analyze the luminescence spectra. All of the experiments were performed in the liquid-helium with a superconducting magnet at a base temperature of $T=4.2$~K. The magnetic field was directed perpendicular to the sample surface and swept in the range of $(0-0.5)$~T. 

\begin{figure}[!t]
\includegraphics[width=\linewidth]{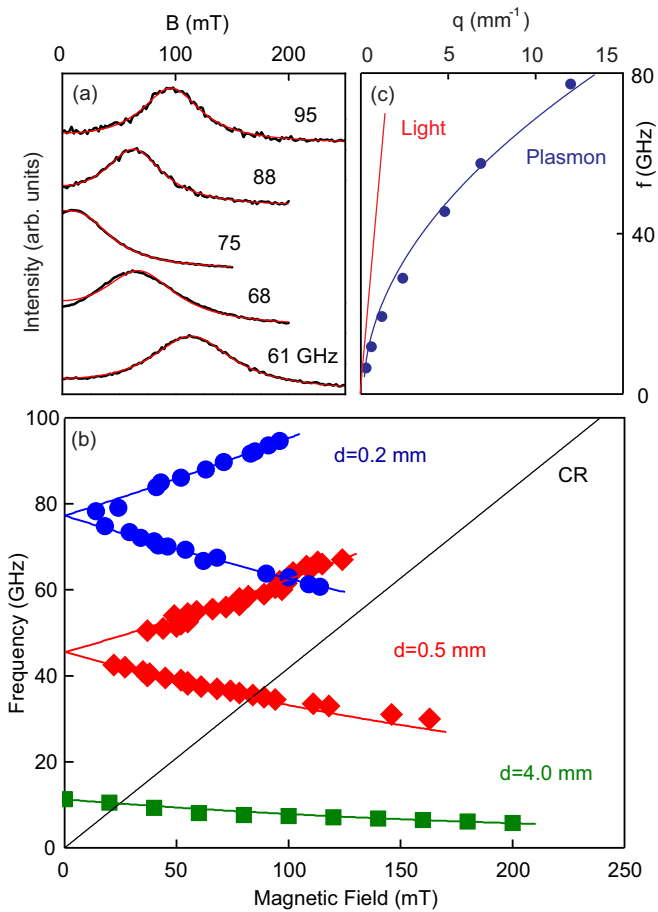}
\caption{(a)~Evolution of the microwave absorption resonances with the excitation frequency for the 2DES disk with a diameter of $d=0.2$~mm and electron density of $n_s=6\times10^{11}$~cm$^{-2}$, the frequency is in the range of $60$~GHz to $100$~GHz. The curves are offset vertically for clarity. (b)~Magnetic-field dependences of the plasmon polariton frequency for the 2DES disks with different diameters. (c)~Dispersion of the hybrid plasmon polariton excitation. The solid lines represent the theoretical prediction of 2D plasmon polariton spectrum and dispersion of the light $\omega_{\rm light} = c q$.} 
\label{fig1}
\end{figure} 

Figure~1(a) shows the experimental magnetic-field dependences of the microwave absorption at different frequencies in a 2DES disk with a diameter of~$d=0.2$~mm and electron density of~$n_s=6\times 10^{11}$~cm$^{-2}$. For convenience, the curves are shifted vertically. Each of the curves shows resonance occurring at different magnetic-field values. The resonance corresponds to the excitation of the plasma oscillation in the disk. In order to obtain the resonance position and half-width, each magneto-absorption curve was fitted using a theory that considers both resonant heating and non-resonant Drude absorption (solid lines in Fig.~1(a)). The detailed fitting procedure is described in the Supplemental Material~\textrm{II}~\cite{Supplemental}. Figure~1(b) shows the magnetic-field values at which plasmon resonances occur as a function of the the microwave frequency $f$ for the disks with diameters of $0.2$, $0.5$ and $4.0$~mm. The theoretical prediction for the magneto-dispersion of plasma waves in the 2DES disks is~\cite{Allen:83, Shikin}:
\begin{equation}
 \omega=\sqrt{\omega_p^2+\left(\frac{\omega_c}{2}\right)^2}\pm\frac{\omega_c}{2},
 \label{2}
\end{equation}
where $\omega_c=eB/m^{\ast}$ is the cyclotron frequency. In the presence of the retardation effects this dependency is modified~\cite{Kukushkin:03, Mikhailov:05, Muravev:06}. 
Both zero-field plasmon frequency and slope of the magnetodispersion dependency decrease. By fitting the theoretical prediction to the data (solid curves in Fig.~1(b)), we can obtain the zero-field plasmon frequency. For the disk under consideration ($d=0.2$~mm), the zero magnetic field plasmon frequency $f_p(0)=78$~GHz. We assign to each plasmon mode a wave vector $q=2.4/d$~\cite{Kukushkin:03}. The experiments of Fig.~1(a) were performed for a serious of 2DES disks with diameters of $d=0.05$, $0.1$, $0.2$, $0.35$, $0.5$, $1$, $2$, $4$ and $8$~mm. The resultant long-wavelength part of the plasmon polariton dispersion at $B=0$~T is plotted in Fig.~1(c). The solid curve represents the theoretical 2D plasmon polariton dispersion~\cite{Kukushkin:03}. The dispersion of light is also shown in the same figure. The plasmon polariton dispersion strongly deviates from the quasi-static $\sqrt{q}$ law, and asymptotically tends to the light line $\omega=cq$ in the limit of a large retardation. 

\begin{figure}[!t]
\includegraphics[width=\linewidth]{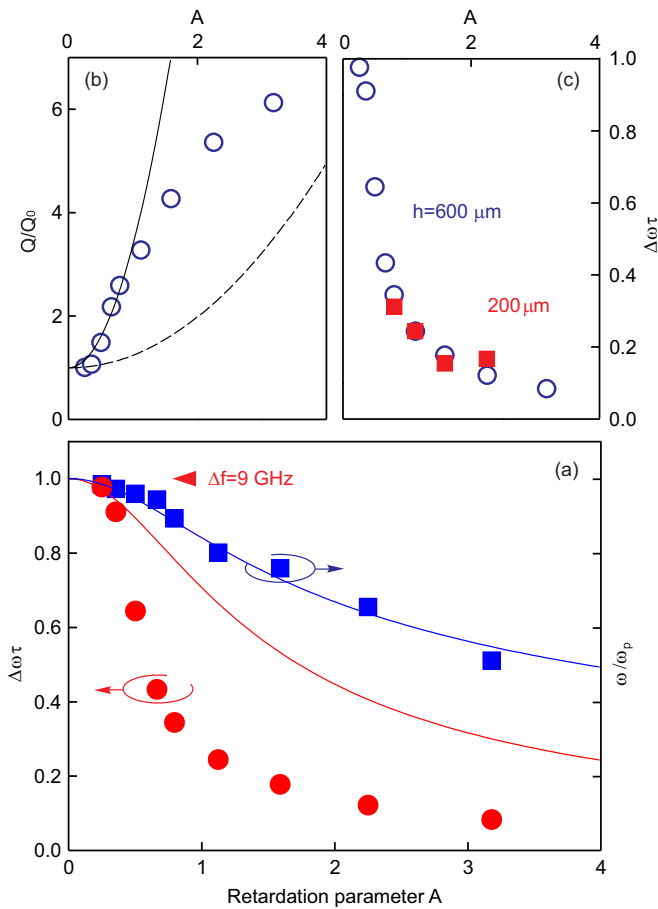}
\caption{(a)~Normalized plasmon damping $\Delta\omega\tau$ (red circles) and frequency $\omega_p/\omega_0$ (blue squares) as a function of the retardation parameter $A$. The solid lines represent the theoretical prediction based on Eq.~(\ref{SSS}). (b)~Dependence of the normalized Q-factor, $Q/Q_0$, of the plasmon polariton excitation on the retardation parameter $A$. The curves represent the quadratic fitting to the experimental points (solid) and theoretical prediction based on Eq.~(\ref{basic}) (dashed). (c)~Comparison of the normalized plasmon half-widths $\Delta\omega\tau$ measured for the samples with thicknesses of $h=0.6$~mm (open circles) and $h=0.2$~mm (filled squares).}
\label{fig2}
\end{figure} 

The red circles in Fig.~2(a) represent the dependence of the normalized plasmon polariton half-width~$\Delta\omega\tau$ on the retardation parameter~$A$. The retardation parameter is calculated using the definition of Eq.~\eqref{1}. The resonance half-width was determined by two mutually consistent methods. The first method directly utilizes frequency scans at $B=0$~T (see inset in Fig.~3).
The second method uses magnetic-field scans at a fixed microwave frequency. The polariton half-width $\Delta \omega$ is then obtained from the half-width of the polariton resonance in $B$ sweeps multiplied by the magneto-dispersion slope $\Delta \omega = (\partial \omega / \partial B) \Delta B$. The $\Delta B$ was obtained by fitting the experimental absorption curves using the theory that considers both resonant heating and non-resonant Drude absorption (see Supplemental Material~\textrm{II}~\cite{Supplemental}). We verified that both methods yield the same value of $\Delta \omega$ for a series of studied disks.

Figure~2(a) also shows the normalized plasmon frequency $\omega/\omega_p$ as a function of $A$ (blue squares). It is remarkable that the plasmon half-width $\Delta\omega\tau$ drastically decreases with $A$. Moreover, the $\Delta\omega\tau$ decreases significantly faster than the resonance frequency $\omega/\omega_p$. These result implies that the retarded plasmon field is concentrated in a region $\lambda_z$, significantly larger than the $\lambda = 2 \pi /q$ region in which the dissipation occurs~\cite{Falko:89}. Indeed, the plasmon polariton damping can be estimated by the formula (we present a precise theoretical derivation in Supplemental Material~\textrm{III}~\cite{Supplemental}):
\begin{equation}
\Delta \omega = \frac{1}{\tau} \times \frac{\lambda}{\lambda_z}.
\label{Purcell}
\end{equation}
The delocalization of the plasmon mode perpendicular to the 2DES plane is determined by $q_z=\sqrt{q^2 - \omega^2/c^2}$. This relation is a direct consequence of the Maxwell's equations for the bound electromagnetic wave propagating along the 2DES, $\Delta \, \vec{E} = (\omega^2/c^2) \, \vec{E}$. Considering the retardation effects, the plasmon polariton dispersion can be very simply expressed as~\cite{Stern:67}:
\begin{equation}
q^2=\frac{\omega^2}{c^2} + \left( \frac{\omega^2}{n_s e^2/2 m^{\ast} \varepsilon_0}  \right)^2.
\label{Dispersion}
\end{equation}
By combining Eq.~(\ref{Purcell}) and Eq.~(\ref{Dispersion}), we can derive the following expressions for the plasmon polariton frequency and damping:
\begin{equation}
\omega^2= \frac{\omega_p^2}{\sqrt{1+A^2}}, \qquad \Delta \omega = \frac{1}{\tau} \frac{1}{\sqrt{1+A^2}}.
\label{SSS}
\end{equation}
In the limit of a not-large retardation, these equations can be expanded in series, leading to:
\begin{equation}
Q=\frac{\omega}{\Delta \omega} = \omega_p \tau \, (1+ A^2)^{1/4} \approx Q_0 \, (1+ A^2/4).
\label{basic}
\end{equation}
We denoted the non-retarded plasmon resonance Q-factor as $Q_0=\omega_p \tau$. It should be noted that Eq.~(\ref{Purcell}) is similar to suppression of the atomic spontaneous emission occurring in the resonator with a unitary Q-factor. This effect has been originally predicted by Purcell~\cite{Purcell}. The coincidence is not accidental, it stems from the unified nature of all electrodynamic phenomena.

Formula~(\ref{SSS}) predicts that $\Delta \omega \tau$ decreases with $A$ faster than the frequency $\omega/\omega_p$. This is consistent with our experimental results in Fig.~2(a). However, although the plasmon polariton frequency closely follows the theory (blue line in Fig.~2(a)), the resonance narrows with the increase of the retardation parameter significantly faster than the theoretical prediction (red line in Fig.~2(a)). In order to establish a direct comparison with our  theory, we plot the normalized Q-factor,~$Q/Q_0$, as a function of the retardation parameter $A$ (Fig.~2(b)). The solid line in Fig.~2(b) represents a quadratic-function $1+kA^2$ ($k=2.4$) fitting. The experimental data closely follow the quadratic dependence for a relatively small retardation ($A<1$). This quantitatively agrees with Eq.~(\ref{basic}), albeit the plasmon damping is suppressed relative to the theory (dashed line in Fig.~2(b)). One may assume that suppression of the plasmon polariton damping is associated with an effect of the semiconductor substrate. In order to test this possibility, we performed measurements (Fig.~2) on the same set of samples thinned from the original $h=0.6$~mm to $h=0.2$~mm. Figure~2(c) shows a comparison of the normalized plasmon half-widths $\Delta\omega\tau$ measured for the samples with thicknesses of $h=0.6$~mm (open circles) and $h=0.2$~mm (filled squares). The main observation to note is that the substrate thickness does not affect the plasmon polariton damping. We can speculate that a more adequate theoretical consideration for the retardation in the studied 2DES geometry is certainly imperative. The discovered phenomenon is significant in terms of both physics and technology. 

\begin{figure}[!t]
\includegraphics[width=\linewidth]{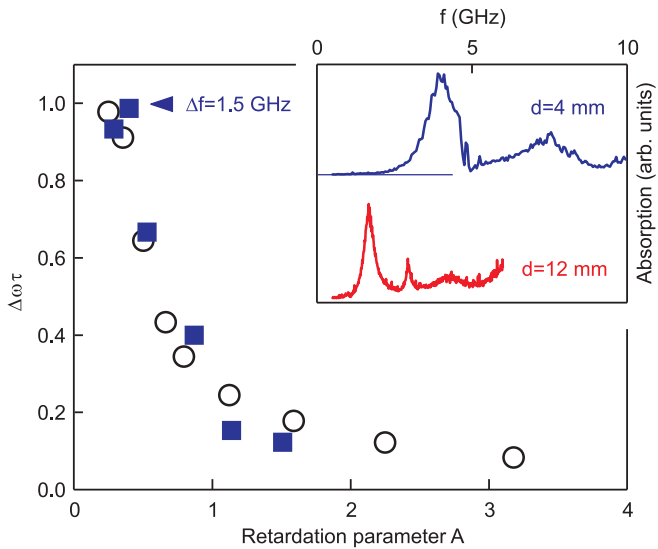}
\caption{Normalized resonance half-width at $B=0$~T as a function of the retardation parameter $A$. The empty circles correspond to the measurements performed on a wafer with $n_s=6\times 10^{11}$~cm$^{-2}$ and $1/\tau = 5.6\times 10^{10}$~s$^{-1}$. The blue squares represent the samples fabricated from different wafers with $n_s=0.8\times 10^{11}$~cm$^{-2}$ and $1/\tau = 9.5\times 10^{9}$~s$^{-1}$. Both dependences are identical. The inset shows microwave absorption spectra measured directly by the frequency scans at $B=0$~T for the 2DES disks with diameters of $d=4$~mm and $d=12$~mm.} 
\label{fig3}
\end{figure} 

One of the most important predictions of the theory is that there is a universal relationship between the normalized plasmon damping  $\Delta \omega \tau$ and retardation parameter $A$. In order to investigate this relation, we repeated the experiment of Fig.~2 for 2DES disks fabricated from three different wafers in terms of density $n_s$ and relaxation time $\tau$. Figure~3 shows the normalized plasmon polariton half-width $\Delta \omega \tau$ as a function of the retardation parameter $A$. The measurements were performed for two wafers with electron densities and relaxations of $n_s=6\times 10^{11}$~cm$^{-2}$ and $1/\tau = 5.6\times 10^{10}$~s$^{-1}$ (empty circles), and $n_s=0.8\times 10^{11}$~cm$^{-2}$ and $1/\tau = 9.5\times 10^{9}$~s$^{-1}$ (blue squares), respectively. The plasmon normalized dampings for the samples fabricated from the two wafers indeed have an idendical dependence on the parameter $A$. The same universal dependence was also reproduced for the third wafer with $n=3.9\times 10^{11}$~cm$^{-2}$ and $1/\tau = 3.4\times 10^{10}$~s$^{-1}$ (see Supplemental Material~\textrm{IV}~\cite{Supplemental}). The discovered universality has a remarkably general nature and is applicable to any type of 2D plasmons. 

In conclusion, we have investigated resonant microwave absorption in 2DES disks with diamaters in the range of $d=0.05$~mm to $12$~mm. The plasmon polariton mode excited in the disks has been detected by the optical technique. We have demonstrated an unexpected anomalous suppression of the plasmon damping when the retardation became important. Remarkably, the dependence of the normalized resonance half-width $\Delta\omega\tau$ on the retardation parameter $A$ has a universal character and does not depend on the 2DES parameters and semiconductor substrate thickness.

We thank L.S.~Levitov and V.A.~Volkov for the stimulating discussions. The authors gratefully acknowledge the financial support from the Russian Science Foundation (Grant No.~14-12-00693).


\begin{thebibliography}{14}

\bibitem{Heitmann:86}
D.~Heitmann, Surf.~Science {\bfseries{170}}, 332 (1986).

\bibitem{Lusakowski:16}
J.\,{\L}usakowski, Semicond.~Sci.~and~Technol. {\bfseries{32}}, 013004 (2016).

\bibitem{Kukushkin:03}
I.~V.~Kukushkin, J.~H.~Smet, S.~A.~Mikhailov, D.~V.~Kulakovskii, K.~von~Klitzing, and W.~Wegscheider, Phys.~Rev.~Lett. {\bf 90}, 156801 (2003).

\bibitem{Muravev:06}
I.~V.~Kukushkin, V.~M.~Muravev, J.~H.~Smet, M.~Hauser, W.~Dietsche, and K.~von~Klitzing, Phys.~Rev.~B {\bf 73}, 113310 (2006).

\bibitem{Stern:67} 
F.~Stern, Phys.~Rev.~Lett. {\bf 18}, 546 (1967).

\bibitem{Gunter:09}
G.~G\"{u}nter, A.~A.~Anappara, J.~Hees, A.~Sell, G.~Biasiol, L.~Sorba,
S.~De~Liberato, C.~Ciuti, A.~Tredicucci, A.~Leitenstorfer, and
R.~Huber, Nature (London) {\bf 458}, 178 (2009).

\bibitem{Todorov:10}
Y.~Todorov, A.~M.~Andrews, R.~Colombelli, S.~De~Liberato, C.~Ciuti, P.~Klang, G.~Strasser, and C.~Sirtori, Phys.~Rev.~Lett. {\bf 105}, 196402 (2010).

\bibitem{Muravev:11}
V.~M.~Muravev, I.~V.~Andreev, I.~V.~Kukushkin, S.~Schmult, and W.~Dietsche, Phys.~Rev.~B {\bf 83}, 075309 (2011).

\bibitem{Science:12}
G.~Scalari, C.~Maissen, D.~Turcinkova, D.~Hagenm\"{u}ller, S.~De~Liberato, C.~Ciuti, C.~Reichl, D.~Schuh,
W.~Wegscheider, M.~Beck, and J.~Faist, Science {\bf 335}, 1323 (2012).

\bibitem{Zhang:16}
Qi~Zhang, M.~Lou, X.~Li, J.~L.~Reno, W.~Pan, J.~D.~Watson, M.~J.~Manfra and J.~Kono, Nature~Physics {\bf 12}, 1005 (2016).


\bibitem{Zudov:14}
Qi~Zhang, T.~Arikawa, E.~Kato, J.~L.~Reno, W.~Pan, J.~D.~Watson, M.~J.~Manfra, M.~A.~Zudov, M.~Tokman, M.~Erukhimova, A.~Belyanin, and J.~Kono,
Phys.~Rev.~Lett. {\bf 113}, 047601 (2014). 

\bibitem{Sirtori:15}
T.~Laurent, Y.~Todorov, A.~Vasanelli, A.~Delteil, C.~Sirtori, I.~Sagnes, and G.~Beaudoin, Phys.~Rev.~Lett. {\bf 115}, 187402 (2015).

\bibitem{Muravev:16}
V.~M.~Muravev, I.~V.~Andreev, S.~I.~Gubarev, V.~N.~Belyanin, and I.~V.~Kukushkin, Phys.~Rev.~B {\bf 93}, 041110(R) (2016).


\bibitem{Sonnichsen:02}
C.~S\"{o}nnichsen, T.~Franzl, T.~Wilk, G.~von~Plessen, J.~Feldmann, O.~Wilson, and P.~Mulvaney, Phys.~Rev.~Lett. {\bf 88}, 077402-1 (2002).

\bibitem{Mikhailov:05}
S.~A.~Mikhailov and N.~A.~Savostianova, Phys.~Rev.~B {\bf 71}, 035320 (2005). 


\bibitem{Andreev:14}
I.~V.~Andreev, V.~M.~Muravev, V.~N.~Belyanin, and I.~V.~Kukushkin, Appl.~Phys.~Lett. {\bfseries{105}}, 202106 (2014).

\bibitem{Shaner}
E.~A.~Shaner, Mark~Lee, M.~C.~Wanke, A.~D.~Grine, J.~L.~Reno, and S.~J.~Allen, Appl.~Phys.~Lett. {\bf 87}, 193507 (2005).

\bibitem{Knap} 
W.~Knap, M.~Dyakonov, D.~Coquillat, F.~Teppe, N.~Dyakonova, J.~{\L}usakowski, K.~Karpierz, M.~Sakowicz, G.~Valusis, D.~Seliuta, I.~Kasalynas, A.~Fatimy, Y.~M.~Meziani, T.~Otsuji, Journal of infrared millimeter and terahertz waves {\bfseries{30}}, 1319 (2009).

\bibitem{Popov}
V.~V.~Popov, D.~V.~Fateev, T.~Otsuji, Y.~M.~Meziani, D.~Coquillat, W.~Knap, Appl.~Phys.~Lett. {\bf 99}, 243504 (2011).


\bibitem{Ashkinadze}
B.\,M.\,Ashkinadze, E.\,Linder, E.\,Cohen, and Arza Ron, Phys. Stat. Sol. {\bf 164}, 231 (1997).

\bibitem{Kukushkin:02}
I.\,V.\,Kukushkin, J.\,H.\,Smet, K.\,von~Klitzing, W.\,Wegscheider, Nature (London) {\bf 415}, 409 (2002).

\bibitem{Supplemental}
See Supplemental Material at XXX, which includes the detailed information about analysis of the plasmon resonance line shape, theory of plasmon polariton spectrum and damping, and additional data on damping of plasmon polariton waves for the third wafer.

\bibitem{Falko:89}
V.~I.~Fal'ko and D.~E.~Khmel'nitskii, Zh.~Eksp.~Teor.~Fiz. {\bfseries{95}}, 1988 (1989) [Sov.~Phys.~JETP {\bfseries{68}}, 1150–1152 (1989)].

\bibitem{Purcell}
E.~M.~Purcell, Phys.~Rev. {\bfseries{69}}, 681 (1946).


\bibitem{Allen:83}
S.~J.~Allen, Jr., H.~L.~St\"{o}rmer, J.~C.~M.~Hwang, Phys.~Rev.~B. {\bf 28}, 4875 (1983).

\bibitem{Shikin}
S.~S.~Nazin and V.~B.~Shikin, Fiz.~Nizk.~Temp. {\bf 15}, 227 (1989) [Sov. J.~Low Temp. Phys. {\bf 15}, 127 (1989)].










\end{thebibliography}
\end{document}


\title{Supplementary Material for\\ ``Observation of universal relation between damping of plasmon polariton waves and the retardation parameter''}
\author{P.~A.~Gusikhin, V.~M.~Muravev,  A.~A.~Zagitova, I.~V.~Kukushkin}
\affiliation{Institute of Solid State Physics, RAS, Chernogolovka, 142432 Russia}

\date{\today}\maketitle

\section{\textrm{I}. Experimental scheme}

\begin{figure}[h!]

\includegraphics[scale=1]{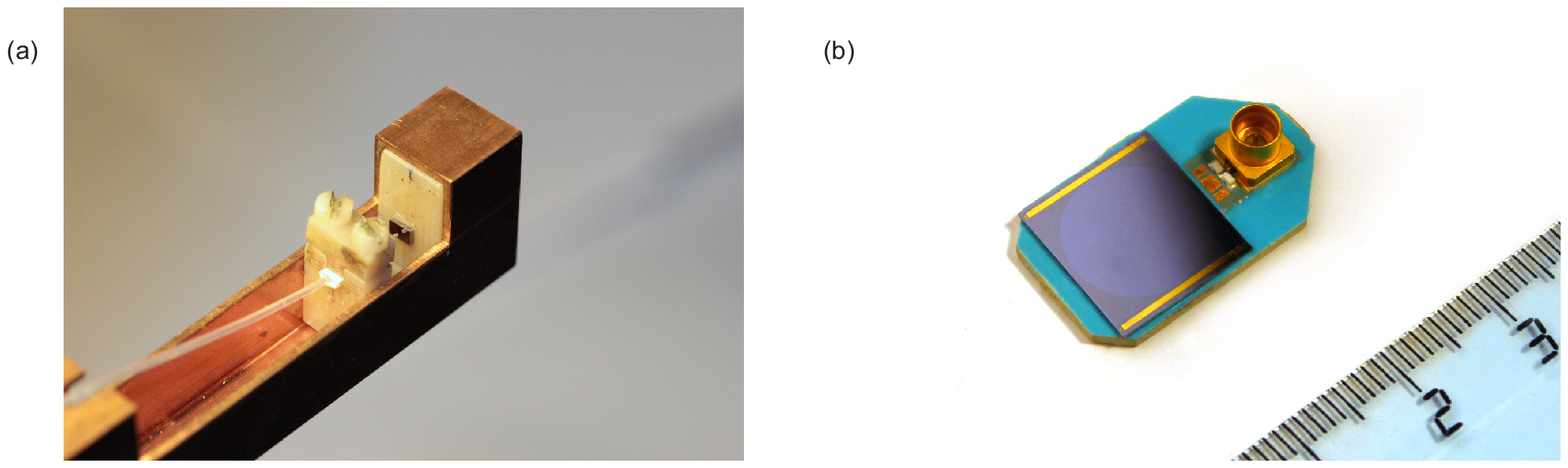}
\caption{(a) Sample placed inside the rectangular waveguide. (b) Sample with $12$~mm 2DES disk and a slot line.}
\label{layout}
\end{figure}

To couple electromagnetic radiation to the disks over a wide range of dimensions, two complementary experimental schemes have been implemented.
In the first waveguide scheme, the sample was placed inside the rectangular waveguide with internal dimensions $15.8$~mm~$\times 7.9$~mm (Fig.~\ref{layout}(a)). 
The sample consists of a single disk. Photo of Fig.~\ref{layout}S(a) shows part of the waveguide with a detached cap in order to mount the sample. During the experiments the cap was attached back.  
The waveguide scheme has been employed for the samples with a diameter of $d\le 4$~mm, and in the frequency range above $f>10$~GHz.

In the second scheme, a single 2DES disk has been fabricated between two metallic gates, which formed a $50~\Omega$ slot line. The sample has been placed on an impedance matched chip-carrier, which was in turn attached to a coaxial cable via SMP connector~(see Fig.~\ref{layout}S(b)). In this way, electromagnetic radiation was deliverd to the 2DES disk. This slot line scheme was used for the samples with disk diameter $d\ge 1$~mm, and in the frequency range $f<40$~GHz.

\section{\textrm{II}. Analysis of the plasmon resonance line shape}

To observe plasma resonances we used optical detection technique based on high sensitivity of the 2DES luminescence spectra to heating caused by the absorption of incident microwave radiation. This technique uses the integral of the absolute value of the differential luminescence spectrum with and without microwave radiation as a measure of a microwave absorption by the 2DES. 
To perform a theoretical analysis of the microwave absorption by 2D plasma, let us consider the motion of a carrier with the isotropic effective mass $m^{*}$ and relaxation time $\tau$ in static magnetic field $B$ perpendicular to the 2DES~\cite{Dresselhaus}:

\begin{figure}[t!]

\includegraphics[scale=1]{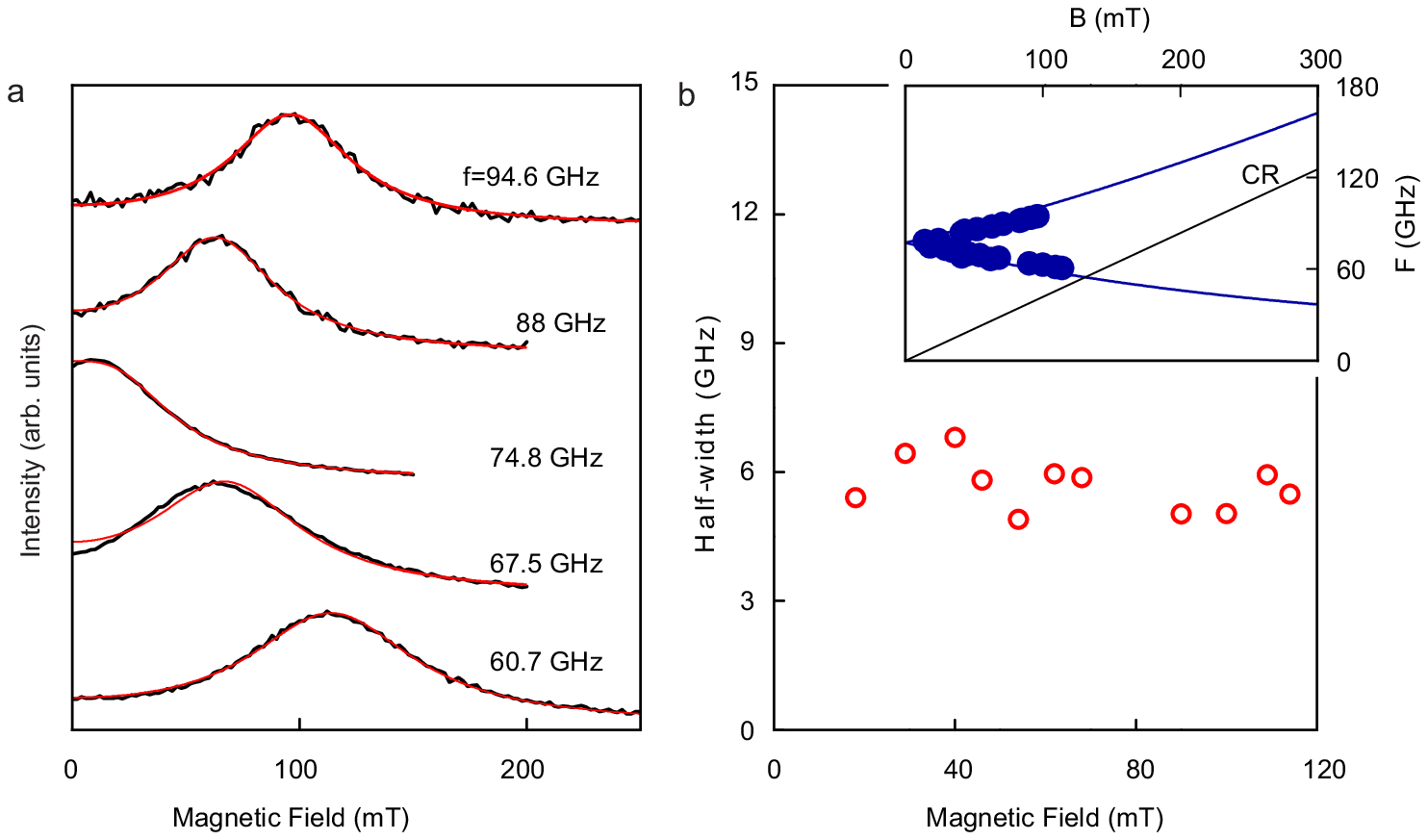}
\caption{(a) Magnetoplasma resonances fitting based on Formula~(\ref{fit2}) for 2DES disk with the electron density  $n_s = 6\times 10^{11}\,\text{cm}^{-2}$ and diameter $d=0.2$~mm. (b)  Plasmon mode half-width versus the resonance magnetic field. The inset shows the magnetodispersion of the plasmon polariton mode.}
\label{image1}
\end{figure}

\begin{equation}
\label{fit}
m^{*}\left(\dfrac{d\upsilon}{dt}+\dfrac{1}{\tau}\upsilon\right)=e\left(E+ \upsilon \times B \right).
\end{equation} 

For plane-polarized radiation $E_{x}$:

\begin{align}
\label{eqn:eqlabel}
\begin{split}
m^{*}\left(i\omega+\dfrac{1}{\tau}\right)\upsilon_{x}=eE_{x} + e \upsilon_{y} B;
\\
m^{*}\left(i\omega+\dfrac{1}{\tau}\right)\upsilon_{y}=-eE_{x}+ e \upsilon_{x} B.
\end{split}
\end{align} 

Then we find for the complex conductivity:

\begin{equation}
\label{fit1}
\sigma=j/E=\sigma_{0}\left(\dfrac{1+i\omega\tau}{1+(\omega_{c}^{2}-\omega^{2})\tau^{2}+2i\omega\tau}\right),
\end{equation}

where $\sigma_{0}=n_s e^{2}\tau/m^{*}$ is the static conductivity of the 2DES with the carrier concentration of $n_s$, $\omega_c = e B/m^{\ast}$ is the cyclotron frequency.
The real part of the conductivity is responsible for the microwave absorption in the 2DES.

\begin{equation}
\label{fit2}
\text{Re }\sigma=\sigma_{0}\left(\dfrac{1+(\omega\tau)^2+(\omega_{c}\tau)^2}{\left(1+(\omega_{c}\tau)^2-(\omega\tau)^2\right)^2+4(\omega\tau)^2}\right)
\end{equation}  

Formula~(\ref{fit2}) was used to fit experimental curves of microwave absorption (red curves in Fig.~\ref{image1}S(a)). There is a remarkable agreement between experiment and theory. The resonance half-width was determined by two mutually consistent methods. The first method directly utilizes frequency scans at $B=0$~T. The second method uses magnetic field scans at a fixed microwave frequency. The polariton half-width $\Delta \omega$ is then obtained from the half-width of the polariton resonance in $B$ sweeps multiplied by the magnetodispersion slope $\Delta \omega = \partial \omega / \partial B \Delta B$. The $\Delta B$ was obtained using fitting of the experimental absorption curves to the theory of Eq.~(\ref{fit2}). We verified that both methods give the same value of $\Delta \omega$ for a series of disks under study. As an example, Figure~\ref{image1}S(b) illustrates this procedure done for the 2DES disk of diameter $d=0.2$~mm and electron density of $n_s = 6\times 10^{11}\,\text{cm}^{-2}$. The extrapolated half-width of the resonance at $B=0$~T is found to be  $\Delta f= (5.7 \pm 0.4)$ GHz. 

\begin{figure}[t!]
\includegraphics[scale=1]{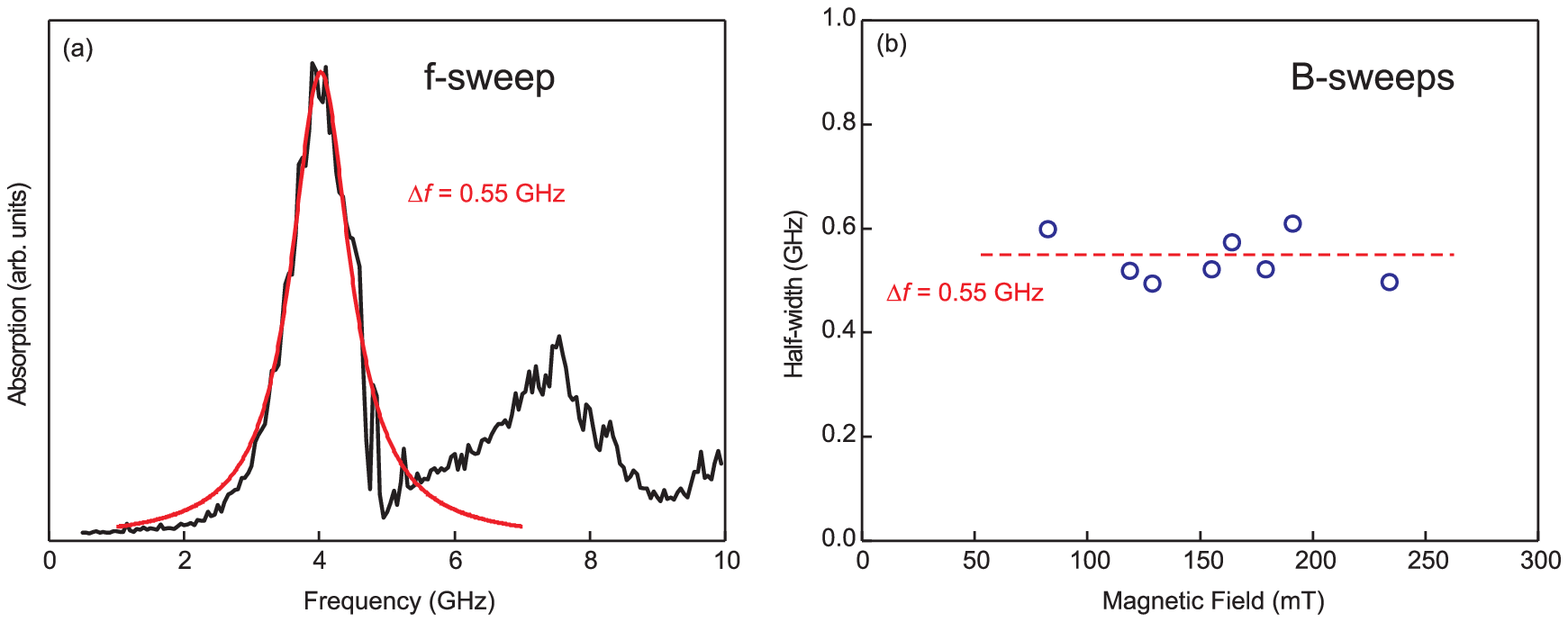}
\caption{(a) Absorption spectrum for 2DES disk with the electron density $n_s = 6\times 10^{11}\,\text{cm}^{-2}$ and diameter $d=4$~mm at zero magnetic field. Red line shows a Lorentzian fitting of the plasma resonance. (b) Magnetic field dependence of the plasmon mode half-width for the same disk calculated from the $B$ sweeps. Red dashed line shows the value $\Delta f=0.55$~GHz.}
\label{comparison}
\end{figure}

To further corroborate that both methods yield the same value of $\Delta\omega$, results of these two methods are summarized in Figure~\ref{comparison}S for the 2DES disk of diameter $d=4$~mm. Figure~\ref{comparison}S(a) shows a direct measurement of the frequency scan at $B = 0$~T. Red curve represents the Lorentzian fitting of the plasmon resonance in the absorption spectrum of the 2DES disk. The resonance half-width $\Delta f=0.55$~GHz is obtained from this fitting. Figure~\ref{comparison}S(b) shows the magnetic field dependence of the plasmon half-width for the same disk. The half-width has been calculated from the $B$ sweeps in the same way, as it was demonstrated above. As one can see, both methods give the same value of $\Delta f= (0.55 \pm 0.05)$~GHz, within the margin of error.

\section{\textrm{III}. Theory of plasmon polariton spectrum and damping}

In the present Supplemental Material we develop theory of plasmon-polariton modes in the infinite two-dimensional electron system located in the vacuum. Let the 2DES is placed in a $z=0$ plane. We start with the set of Maxwell equations for the vector and scalar potentials $(\mathbf{A},A_z)$ and $\varphi$~\cite{Falko:89}:

\begin{equation}
\left(\frac{1}{c^2}\frac{\partial^2}{\partial t^2}-\Delta\right)\left\{
\begin{array}{ccc} \varphi\\ A_z\\ \mathbf{A} \end{array}
\right\}=\left\{
\begin{array}{ccc} \rho/\varepsilon_0\\ 0\\ \mu_0\mathbf{j} \end{array}
\right\}\delta(z),
\label{Max:1}
\end{equation}

\begin{equation}
\nabla\mathbf{A}+\frac{\partial A_z}{\partial z}+\frac{1}{c}\frac{\partial\varphi}{\partial t}=0.
\label{Max:2}
\end{equation}

We seek the oscillations in the system in the form of a wave $\exp(i\mathbf{q}\mathbf{r}-i\omega t)$. We assume that the field of the wave is localized near the $z=0$ plane. Therefore, from equations~\eqref{Max:1}~and~\eqref{Max:2} we obtain that the potentials $\mathbf{A}$, $A_z$ and $\varphi$ are proportional to $\exp(-\kappa|z|)$ where
$$
\kappa=q_z=\sqrt{q^2-\left(\frac{\omega}{c}\right)^2}.
$$

Using~\eqref{Max:1},~\eqref{Max:2} and a material equation:
\begin{equation}
\mathbf{j}=-\sigma\left(\nabla\varphi+\frac{1}{c}\frac{\partial\mathbf{A}}{\partial t}\right)
\label{Max:3}
\end{equation}
we get the dispersion equation for a longitudinal waves ($\mathbf{A}\parallel\mathbf{q}$):

\begin{equation}
i\omega=\frac{\sigma}{2\varepsilon_0}\sqrt{q^2-\left(\frac{\omega}{c}\right)^2},
\label{eq:1}
\end{equation}
where $\sigma$ is a frequency dependent 2D conductivity:
\begin{equation}
\sigma(\omega)=\sigma_0\frac{\gamma}{\gamma-i\omega}.
\label{eq:2}
\end{equation}
Here, $\gamma=1/\tau$. Let $x=\sigma_0/(2\varepsilon_0 c)$, then \eqref{eq:1} is transformed to the form:
\begin{equation}
\omega^4+2i\gamma\omega^3+(x^2-1)\gamma^2\omega^2-x^2\gamma^2c^2q^2=0.
\label{eq:3}
\end{equation}
Given $\omega=\omega_1+i\omega_2$, where $\omega_2=-\Delta\omega/2$, we get the system of equations:
\begin{equation}
\begin{cases}
\omega_1^4-6\omega_1^2\omega_2^2+\omega_2^4-6\gamma\omega_1^2\omega_2+2\gamma\omega_2^3+(x^2-1)\gamma^2(\omega_1^2-\omega_2^2)-x^2\gamma^2c^2q^2=0,\\
2\omega_1\left[2\omega_1^2\omega_2-2\omega_2^3+\gamma\omega_1^2-3\gamma\omega_2^2+(x^2-1)\gamma^2\omega_2\right]=0.\\
\end{cases}
\label{eq:4}
\end{equation}
Physically meaningful solutions should satisfy the conditions: $\omega_1>0$ and $\omega_2<0$. In the limit of small wavelengths $q \to\infty$, one can find a well-known 2D plasmon dispersion: 
\begin{equation}
\omega=\sqrt{x\gamma c q}-i\frac{\gamma}{2}.
\label{eq:5}
\end{equation}

To obtain approximate solution for the plasmon polariton spectrum and damping in the retarded limit, we define $\omega_2=-\gamma/2+t$, where $t\to 0$ at $q\to\infty$. From the second equation of the \eqref{eq:4} we get:
\begin{equation}
\omega_1^2=t^2-\frac{(2x^2+1)\gamma^2}{4}+\frac{x^2\gamma^3}{4t}.
\label{eq:6}
\end{equation}
Substituting that to the first equation of the \eqref{eq:4} and maintaining the principal terms:
\begin{equation}
\frac{x^4\gamma^6}{16t^2}-\frac{x^2\gamma^5}{8t}-x^2\gamma^2c^2 q^2\approx 0 \qquad \Rightarrow \qquad t\approx\frac{x\gamma^2}{4cq}-\frac{\gamma^3}{16c^2q^2}.
\label{eq:7}
\end{equation}
From \eqref{eq:6} and \eqref{eq:7}:
\begin{equation}
\omega_1\approx\sqrt{x\gamma cq}\left(1-\frac{x\gamma}{4cq}\right), \qquad \omega_2\approx -\frac{\gamma}{2}\left(1-\frac{x\gamma}{2cq}\right).
\label{eq:8}
\end{equation}
Finally, we obtain for the plasmon polariton Q-factor:
\begin{equation}
Q=-\frac{\omega_1}{2\omega_2}\approx\sqrt{\frac{xcq}{\gamma}}\left(1+\frac{x\gamma}{4cq}\right)=Q_0\left(1+\frac{A^2}{4}\right),
\label{eq:9}
\end{equation}
where $Q_0=\sqrt{xcq/\gamma}=\omega_p \tau$ is the Q-factor of the 2D plasmon, $A=\sqrt{x\gamma/cq}$ is the retardation parameter. Notably, obtained expression for the plasmon polariton Q-factor when the influence of retardation effects is prominent is fully equivalent to the Eq.~(6) of the paper.

\section{\textrm{IV}. Universal relation between damping of plasmon polariton waves and the retardation parameter - Wafer~\textrm{III}}

\begin{figure}[!h]

\includegraphics[scale=1]{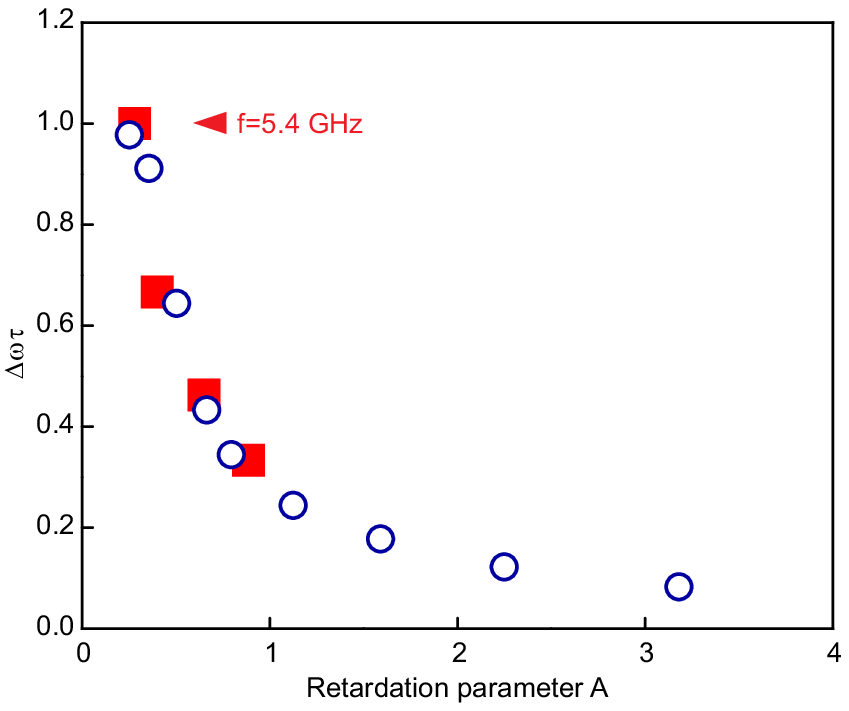}
\caption{Normalized plasmon damping $\Delta \omega \tau$ versus the retardation parameter $A$ measured for the wafer~\textrm{I} (blue circles) and wafer~\textrm{III} (red squares). The dependency has a universal nature independent of the 2DES parameters.}
\label{image3}
\end{figure}

One of the most important predictions of the theory is that normalized plasmon damping  $\Delta \omega \tau$ and retardation parameter $A$ have a universal relationship. In order to confirm this again, we measured the plasmon polariton half-width $\Delta \omega \tau$ as a function of the retardation parameter $A$ for the wafer~\textrm{III} with $n_s = 3.9\times 10^{11}\ \text{cm}^{-2}$ and $1/\tau = 3.4 \times 10^{10}$~s$^{-1}$ (red squares in Fig.~\ref{image3}S). The experiments were performed on 2DES disks with diameters  $d=0.1, \, 0.2, \,  0.6$ and $1$ mm. For comparison, the same plot includes $\Delta \omega \tau$ as a function of $A$ measured for the wafer~\textrm{I} ($n_s=6\times 10^{11}$~cm$^{-2}$, $1/\tau = 5.6\times 10^{10}$~s$^{-1}$). When the retardation effects are negligible the resonance half-width for the wafer with the electron density $n_s = 3.9\times 10^{11}\ \text{cm}^{-2}$ is found to be $\Delta f=5.4$ GHz. As the retardation parameter increases, the normalized half-width $\Delta \omega \tau$ decreases in the same way for both wafers. Therefore, we conclude that the normalized resonance half-width dependence on retardation parameter has a universal nature independent of the 2DES parameters, as previously discussed in the paper. 

